\documentstyle[aps,epsf,twocolumn]{revtex}
\begin{document}
\twocolumn[\hsize\textwidth\columnwidth\hsize\csname @twocolumnfalse\endcsname
 
\title{S-wave Superconductivity in Weak Ferromagnetic Metals }
\author{K.B. Blagoev, J.R. Engelbrecht and K.S. Bedell}
\address{Physics Department, Boston College, Chestnut Hill, MA 02167}
\date{\today}
\maketitle
\begin{abstract}
We investigate the behavior of weak ferromagnetic metals close to the
ferromagnetic critical point. 
We show that in the limit of small magnetic moment
the low temperature metallic phase is rigorously described by a local 
ferromagnetic Fermi liquid that has a momentum-independent self-energy.
Whereas, non-Fermi liquid features develop at higher temperatures.
Furthermore,
we find that an instability towards $s$-wave superconductivity is
possible when the exchange splitting is comparable to the superconducting gap.
\end{abstract}
\pacs{PACS numbers: 71.10.+x, 71.27.+a, 74.10.+v, 75.10.Lp}
]

Itinerant ferromagnetic materials have recently been studied intensively
experimentally and theoretically because of potential applications as well
as their interesting physical properties. The strong electronic correlations
in these materials are important close to the critical point and leads to a
variety of physical behavior. Whereas the colossal magnetoresistive
materials \cite{vonHelmolt&Jin93&94,Shimakawa+Subramanian96,Ramirez et al.97}
exhibit a ferromagnetic-metal to insulator transition at finite
temperatures, weak ferromagnetic materials such as MnSi, ZrZn$_2$\cite
{Lonzarich} and some of the heavy fermion materials\cite{Stewart97} can be
tuned through a ferromagnetic to nonferromagnetic transition in the ground
state. It is near this critical point that the strong electronic
correlations can result in a Fermi liquid to non-Fermi liquid transition.
Although theoretically these systems have been studied intensively\cite
{Sachdev+Continentino,TheorCMR} the role of the electronic correlations
close to the phase transition on the magnetic side is not clear.

When pressure is applied to the weak ferromagnets, MnSi and ZrZn$_2$\cite
{Lonzarich}, the magnetic moment ($m_0$) as well as the Curie temperature ($%
T_{{\it c}}$) are driven to zero. When $T_{{\it c}}$ approaches zero these
materials exhibit features that are characteristic of a system close to a
quantum phase transition. Much of the recent theoretical and experimental
focus\cite{Lonzarich,Sachdev+Continentino} has been on this quantum critical
regime. The physics of weak ferromagnetic metals can be understood within
the framework of the ferromagnetic Fermi liquid (FFL) theory developed first
by Abrikosov and Dzyaloshinskii\cite{Abrikosov&Dzyaloshinskii58}, and
Dzyaloshinskii and Kondratenko\cite{Dzyaloshinskii&Kondratenko76}.

In this letter we discuss several new and unexpected results for weak
ferromagnetic metals. The most dramatic is the strong coupling between spin
and charge fluctuations in the limit of vanishing magnetic moment. This
opens up the possibility of an $s$-wave superconducting instability in weak
ferromagnetic materials. This is clearly counter to much of the early work
linking weak ferromagnetism and $p$-wave superconductivity\cite
{Enz&Matthias78}. The physics responsible for the enhanced $s$-wave and
suppressed $p$-wave pairing is the local nature of the self-energy, $\Sigma
(\epsilon ,p)$. A local self-energy depends only on the frequency. From this
it follows that only the $l=0$ interactions between the quasiparticles
survive\cite{Engelbrecht&Bedell95}. An additional constraint coming from the
Pauli principle leads to the strong coupling between the spin and charge
fluctuations. In the limit of weak ferromagnetism, close to the quantum
critical point the locality of the self-energy becomes rigorous. The paper
is structured as follows. We will first explore the properties of a local
FFL and characterize some of the instabilities when $m_0\rightarrow 0$.
After the discussion of the superconductivity we will outline the
microscopic properties of a local Fermi liquid. We then show that the
electron self-energy becomes local in weak ferromagnets. This will be
followed by a proposed phase-diagram and some discussion of the nature of
the state on the paramagnetic side of the phase diagram.

We start off introducing the Fermi liquid theory for a ferromagnetic metal.
This theory, of course, differs from that of a paramagnetic Fermi liquid, in
that in a ferromagnet a magnetic moment (or internal field) is spontaneously
generated by the interactions. The formation of this state is
nonperturbative, arising from the types of singularities assumed in the
Green's function and interaction vertex. In what follows the question of
which Hamiltonian generates the ferromagnetic state is not relevant. Here we
will assume that the ground state of our system is a ferromagnetic Fermi
liquid.

The deviation of the energy from its equilibrium value can be expanded up to
second order in the deviations, $\delta n_{{\bf p}\sigma }$ of the momentum
distribution function. The form for this is \cite{AGD75,Baym&Pethick91}

\begin{equation}
\delta E=\sum_{{\bf p}\sigma }\varepsilon _{{\bf p}\sigma }^0\delta n_{{\bf p%
}\sigma }+\frac 12\sum_{{\bf p}\sigma ,{\bf p}^{\prime }\sigma ^{\prime }}f_{%
{\bf pp}^{\prime }}^{\sigma \sigma ^{\prime }}\delta n_{{\bf p}\sigma
}\delta n_{{\bf p}^{\prime }\sigma ^{\prime }}+...  \label{one}
\end{equation}

where $\varepsilon _{{\bf p}\sigma }^0$ is the quasiparticle energy, $f_{%
{\bf pp}^{\prime }}^{\sigma \sigma ^{\prime }}=f_{{\bf pp}^{\prime
}}^{\sigma \sigma ,\sigma ^{\prime }\sigma ^{\prime }}$ are the
quasiparticle interactions in the presence of the internal field, with the
volume of the system set to unity, and $\sigma =\uparrow $,$\downarrow $.
This expression, Eq.(\ref{one}), is valid for any value of the magnetic
moment, $m_0$ (where $m_0=n_{\uparrow }-n_{\downarrow }$, $n_\sigma $ is the
occupation number of particles with spin projection $\sigma $, and the
magneton is $\mu =1$). In general\cite{Dzyaloshinskii&Kondratenko76,AGD75},
(neglecting momentum labels) $f^{\sigma \sigma ^{\prime }}=f_s+(\sigma
_x\sigma _x^{\prime }+\sigma _y\sigma _y^{\prime })f_a^I+\sigma _z\sigma
_z^{\prime }f_a^{II}+(\vec{\sigma}\cdot \vec{m}_0)(\vec{\sigma}^{\prime
}\cdot \vec{m}_0)f_m^I+(\vec{\sigma}\cdot \vec{\sigma}^{\prime
})m_0^2f_m^{II}$. For our purpose we are interested in the case when $%
m_0/n\ll 1$ ($n=n_{\uparrow }+n_{\downarrow }$), i.e., the limit of weak
ferromagnetism. In this limit we can treat the quasiparticle interaction as
rotationally invariant in spin space, thus 
\begin{equation}
f_{{\bf pp}^{\prime }}^{\sigma \sigma ^{\prime }}=f_{{\bf pp}^{\prime }}^{%
{\bf s}}+f_{{\bf pp}^{\prime }}^{{\bf a}}{\bf \sigma }\cdot {\bf \sigma }%
^{\prime }+{\cal O}(m_0^2)  \label{two}
\end{equation}
where the superscript $s$ ($a$) stands for the symmetric (antisymmetric)
components. The corrections depending on the magnetization are unimportant
for what follows and will be omitted. The Fermi liquid parameters, $%
F_l^{s,a} $are therefore $N(0)f_{{\bf pp}^{\prime }}^{{\bf \sigma \sigma }%
^{\prime }}=\sum_l(F_l^s+F_l^a{\bf \sigma }\cdot {\bf \sigma }^{\prime })P_l(%
{\bf \hat{p}\cdot \hat{p}}^{\prime })$, where $P_l({\bf \hat{p}\cdot \hat{p}}%
^{\prime })$ are the Legendre polynomials. Similarly, one can expand the
scattering amplitudes $N(0)a_{{\bf pp}^{\prime }}^{{\bf \sigma \sigma }%
^{\prime }}=\sum_l(A_l^s+A_l^a{\bf \sigma }\cdot {\bf \sigma }^{\prime })P_l(%
{\bf \hat{p}\cdot \hat{p}}^{\prime })$.

The local self energy implies that the only non-zero Fermi liquid parameters
are the $l=0$ parameters: $F_0^a(m_0)$ and $F_0^s(m_0)$ \cite
{Engelbrecht&Bedell95}. The Fermi liquid parameters close to the transition
point can be expanded up to second order in the magnetization\cite
{Bedell&Sanchez-Castro86}(the order parameter for the magnetic transition).
Substitution of this expansion in Eq.(\ref{one}) gives 
\begin{equation}
\delta E=\frac{1+F_0^a}{2N(0)}m_0^2+g\frac 1{N(0)^3}m_0^4+...
\end{equation}
where $g$ is a positive constant and the Fermi liquid parameters are
magnetization independent. The minimum of the energy for $F_0^a<-1$, occurs
at the equilibrium magnetization 
\begin{equation}
m_0\sim |1+F_0^a|^{1/2}
\end{equation}
and in the limit $F_0^a\rightarrow -1^{-}$ the equilibrium magnetization
goes to zero. This is the weak FFL. In the case of weak quasiparticle
interaction the Fermi liquid parameters are related to the scattering
coefficients through\cite{Baym&Pethick91} 
\begin{equation}
A_l^\alpha =\frac{F_l^\alpha }{1+F_l^\alpha /(2l+1)}\text{, }\ \alpha =a,s%
\text{.}  \label{scfer}
\end{equation}
Therefore only the $l=0$ scattering amplitudes are nonzero. An important
additional simplification occurs as a consequence of the Pauli principle.
This is the forward scattering sum rule. It states that the triplet
scattering amplitude is zero, i.e. $a^{\uparrow \uparrow }=0$. When we
expand $a^{\uparrow \uparrow }$ in powers of $m_0$, all coefficients must
vanish. The zeroth order term gives 
\begin{equation}
A_0^a+A_0^s=0  \label{Pauli2}
\end{equation}
The next term is second order in $m_0$ and the coefficients involve
derivatives of $A_l^\alpha $, $\alpha =a,s$ etc. Therefore the weak
ferromagnetic Fermi liquid has only two independent parameters: say $A_0^a$
and $m^{*}/m$, where $m$ and $m^{*}$ are the bare and quasiparticle mass
respectively. The Fermi liquid parameters are therefore not independent and
Eqs.(\ref{scfer}) and (\ref{Pauli2}) give 
\begin{equation}
F_0^s=-\frac{F_0^a}{1+2F_0^a}
\end{equation}
In the limit of weak ferromagnetism $F_0^a\rightarrow -1^{-}$ it follows
that $F_0^s\rightarrow -1^{+}$. In this limit the scattering amplitudes are: 
$A_0^a$ $\rightarrow +\infty $ and $A_0^s$ $\rightarrow -\infty $,
indicating an instability in the spin and charge sector respectively. This
leads to phase separation at the point of the magnetic phase transition and
the compressibility in this limit is 
\begin{equation}
\kappa =\frac 1{n^2}\frac{N(0)}{1+F_0^s}\rightarrow \infty .
\end{equation}
Spin and charge are strongly coupled by the Pauli principle (Eq.(\ref{Pauli2}%
)) and the singularity in the compressibility is connected to the
singularity in the susceptibility. The phase transition occurs at the same
point in the spin and charge sector.

The simple physics described above can be altered in the vicinity of the
phase transition and can lead to an $s$-wave superconducting state. When the
exchange splitting - $v_F(p_{\uparrow }-p_{\downarrow })$ becomes comparable
to the superconducting gap, $\Delta $, the Larkin-Ovchinnikov-Fulde-Ferrell%
\cite{LOFF64} state should be favored with a finite total momentum of the
pair. The triplet scattering amplitude is zero and the singlet scattering
amplitude is 
\begin{equation}
N(0)a^{sing}=-4\frac{F_0^a}{1+F_0^a}.
\end{equation}
At finite temperatures one would expect 
\begin{equation}
N(0)a^{sing}\sim \frac{N(0)\lambda }{1+N(0)\lambda [\ln T^{*}/T]}
\end{equation}
where $T^{*}=T_c^2/\epsilon _F\ll T_c$ is a temperature below which our
considerations are valid. Here we do not attempt to calculate the critical
temperature for the superconducting transition, but merely state that such a
temperature exists. Since $a^{sing}$ is negative and the logarithm in the
denominator is positive, $\lambda $ is negative. For $\lambda <0$ an $s$%
-wave pairing state must occur for some temperature less than $T^{*}$.
Although in the paramagnetic phase local spin fluctuations destroy the $s$%
-state pairing, in the weakly ferromagnetic state the charge and spin are
strongly coupled, expressed in the locality of the self-energy, giving rise
to an $s$-wave pairing.

This state should in principle be observable at very low temperatures. To
estimate this low temperature we note that the above calculations, although
done at zero temperature are valid for temperatures $T\ll T^{*}$ where $T_c$
is the ferromagnetic transition temperature and $\epsilon _F$ is the Fermi
energy. This state could be observed in MnSi under pressure where for
example $T_c=30K$ when $P=14kbar$. In this pressure range we expect that $%
T^{*}$ is around $1K$ and the order of magnitude for the superconducting
transition temperature is around $10mK$. This should be possible to achieve
in practice.

A phonon mechanism for the appearance of ferromagnetism in the weak
ferromagnetic metal $ZrZn_2$ has been suggested by Enz and Matthias \cite
{Enz&Matthias78} (later found also in other compounds\cite{Giorgi et al.79}%
). According to their theory the ferromagnetism occurs as a result of
suppressed $p$-wave superconductivity leading to a Stoner instability.
Although it seems natural to assume that the $p$-wave pairing and
ferromagnetic order are closely related and compete in different systems, we
have shown that the Pauli principle necessarily requires $s$-wave pairing.
This is contradictory to the simple picture of a ferromagnetically
suppressed $p$-wave superconductivity.

The self energy in the weak ferromagnetic Fermi liquid, close to the Fermi
surface is weakly momentum dependent and as in the electron-phonon problem
the main contribution comes from the frequency dependence. If we chose the
magnetization axis along the $\hat{z}$ axis, the single particle Green's
function close to the Fermi surface is 
\begin{equation}
G_\sigma (p)=\frac z{\epsilon -v_F(|{\bf p}|-p_\sigma )\pm i\delta }%
+G_\sigma ^{inc}(p)
\end{equation}
where $G_\sigma ^{inc}$ is the nonsingular part of the Green's function, $%
p=(\varepsilon ,{\bf p})$ is the energy-momentum vector, $p_\sigma $ is the
Fermi momentum for particles with the given spin orientation, $v_F$ is the
Fermi velocity, and $z$ is the quasiparticle residue. In the vicinity of the
phase transition, the small parameter due to the exchange splitting is $%
\theta =p_{\uparrow }-p_{\downarrow }\ll p_{\uparrow }$,$p_{\downarrow }$.
The velocity difference at the two Fermi surfaces and the corresponding
residues give correction to the quasiparticle energy difference of the order 
${\it O}(\theta ^2/p_s^2)$ and are ignored. The liquid has two types of low
energy collective spin excitations\cite{Dzyaloshinskii&Kondratenko76}. The
longitudinal spin fluctuations are paramagnons while the transverse are spin
waves. The self energy (similarly to the approximation used in the
electron-phonon problem\cite{Schrieffer64}) is 
\begin{equation}
\Sigma _\sigma (p)\simeq \frac{\tilde{g}^2}i\!\!\int \!\!\frac{d^4q}{(2\pi
)^4}[G_\sigma (p\!+\!q)\chi _{\parallel }(q)\!+2\!G_{\bar{\sigma}%
}(p\!+\!q)\chi _{\pm }(q)]
\end{equation}
where $\bar{\sigma}$ is the opposite to $\sigma $ , $\tilde{g}$ is an
effective coupling constant for the spin-spin interaction between the
quasiparticles. The contribution to the second term in the expression for
the self-energy comes from values of the momentum corresponding to the
maximum frequency of the spin waves. Therefore the relevant expressions for
the spin susceptibilities are given by\cite{Dzyaloshinskii&Kondratenko76} 
\begin{equation}
\chi _{\parallel }(k)=\frac{N(0)}2\frac 1{(\theta /p_F)^2+b^2k^2-i\pi \omega
/2v_F|{\bf k}|}
\end{equation}
\begin{equation}
\chi _{\pm }(k)=\frac{N(0)}2\frac 1{b^2k^2-i\pi \omega /2v_F|{\bf k}|}
\end{equation}
Here $N(0)=m^{*}p_F/2\pi ^2$ is the density of particle states at the Fermi
surface with $N_{\uparrow }(0)\simeq N_{\downarrow }(0)\equiv N(0)$ and $%
b\sim p_F^{-1}$. $p_F$ is the Fermi momentum. A standard change of variables%
\cite{Schrieffer64} and taking into account that we are looking at phenomena
in the vicinity of the Fermi surface we have for the self energy 
\begin{eqnarray}
\Sigma _\sigma (\epsilon ) &\simeq &\frac{i\tilde{g}^2}{(2\pi )^3}%
\int_0^{p_c}\!\!\!qdq\int_{-\infty }^{+\infty }\!\!\!\!d\omega \times  \\
&&\{\int_{p_\sigma \!-\!q}^{p_\sigma \!+\!q}\!\!p^{\prime }{\frac{dp^{\prime
}}{p_\sigma }}\frac{\chi _{\parallel }(q)}{\omega +\epsilon -\xi _{p^{\prime
}}^\sigma +i\delta [p^{\prime }-p_\sigma ]}  \nonumber \\
&&+2\int_{p_{\bar{\sigma}}\!-\!q}^{p_{\bar{\sigma}}\!+\!q}\!\!p^{\prime }{%
\frac{dp^{\prime }}{p_{\bar{\sigma}}}}\frac{\chi _{\pm }(q)}{\omega
+\epsilon -\xi _{p^{\prime }}^{\bar{\sigma}}+i\delta [p^{\prime }-p_{\bar{%
\sigma}}]}\}  \nonumber
\end{eqnarray}
which is momentum independent. Here, $\xi _{p^{\prime }}^\sigma
=v_F(p^{\prime }-p_\sigma )$ and we have gone from the variables ${\bf q}%
=(q_x,q_y,q_z)$ to $(q=|{\bf q}|,p^{\prime }=|{\bf p}+{\bf q}|,\varphi
=\arctan q_y/q_x)$ and $i\delta [q]\equiv i\delta sign(q)$. The real part
the self energy is momentum independent and the effective mass, $m^{*}\equiv
m(1-\frac{\partial \Sigma }{\partial \omega })$, renormalized by the
spin-fluctuations is $\sim -\ln \theta /p_c$ (here $p_c$ is an ultraviolet
cutoff reflecting the unknown large momentum physics) and is divergent as
the system approaches the phase transition from the ferromagnetic side, i.e. 
$m_0\rightarrow 0$. Therefore the quasiparticle residue goes to zero and a
non-Fermi liquid is approached. We expect at small magnetizations
competition between superconductivity and a non-Fermi liquid.

It must be noted that only in the energy interval, $T\ll T^{*}<T_c$ we
expect a Fermi liquid behavior (Fig.1). Up to now we considered the zero
temperature transition point.  This point is approached by varying the
interaction strength. As this point is approached the effective mass,
calculated in the FFL diverges logarithmically indicating a metal insulator
transition. Away fom the quantum critical point there is a regime for $%
T^{*}\ll T<T_c$ in which the spin fluctuations modify the liquid and the
specific heat is\cite{Dzyaloshinskii&Kondratenko76} 
\begin{equation}
C_s\approx \frac T{T_0}\ln \frac{T_0}T
\end{equation}
giving rise to a non-Fermi liquid state which has been observed\cite
{Lonzarich,Stewart97} experimentally. In these experiments the conductivity
increases but remains finite across the magnetic phase transition line.
Therefore the system enters a metallic nonmagnetic regime. However, in the
non Fermi liquid the effective mass, connected to the quasiparticle residue,
is not equal to the optical conductivity mass which remains finite\cite
{Blagoev et al. tbp} and therefore the system stays metallic.

Recently two of us\cite{Engelbrecht&Bedell95} studied a local paramagnetic
Fermi liquid and showed that its ground state is robust against a phase
transition from a paramagnetic metal to a ferromagnetic insulator. This
result, recently, has also been obtained in a local model through a scaling
analysis\cite{Belitz et al.96}. Our opinion is that existence of the phase
transition is due to the non-locality of the self energy in the neighborhood
of the ferromagnetic instability. Other local theories are the dynamical
mean field theories, that have been intensively studied\cite{Georges et
al.96} numerically. Recently, there have been calculations showing that a
ferromagnetic instability occurs in the Hubbard model on an fcc lattice with
a local self energy at an enhanced strength of the interaction compared to
the case of a nonlocal self energy\cite{Herrmann&Nolting97;Ulmke95}. A
probable cause is that in these calculations the paramagnetic state is not a
Fermi liquid and therefore a ferromagnetic instability would be possible.

\begin{figure}[h]
\vspace{0.3cm}
\epsfxsize=7.0cm
\hspace*{0.2cm}
\epsfbox{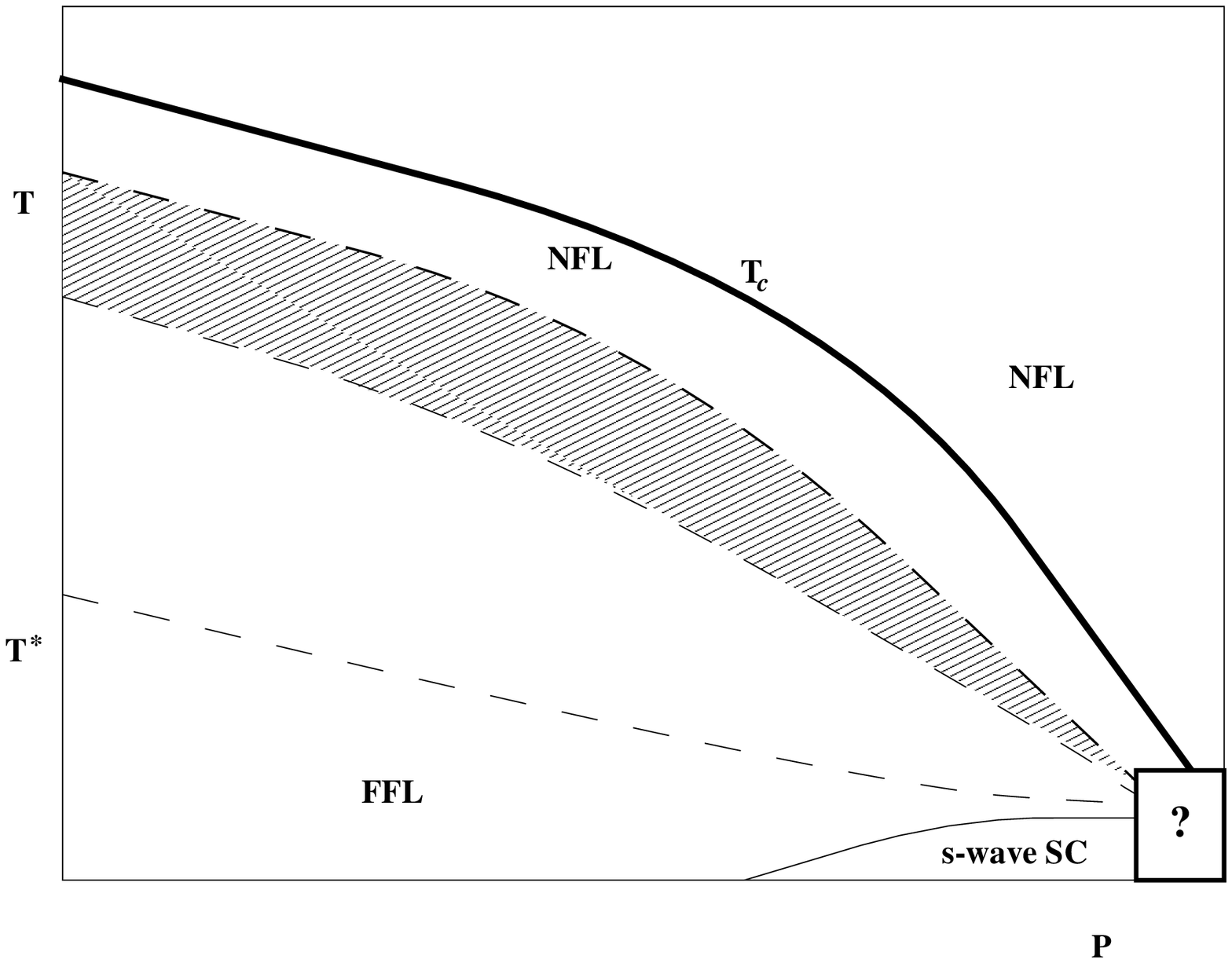}
%\vspace{-5.2cm}
\caption{Schematic phase diagram of a weak ferromagnetic Fermi liquid. The
lines are only guidance to the eye. s-wave SC: s-wave superconducting state;
NFL: non-Fermi liquid state; FFL: ferromagnetic Fermi liquid state}
%\vspace{-0.26cm}
\label{fig}
\end{figure}

We can summarize the above calculations by a schematic phase diagram of a
weak ferromagnetic metal, Fig.(1). The phase transitions from the
ferromagnetic state to the Larkin-Ovchinnikov-Fulde-Ferrell (LOFF) state is
second order while the transition from the LOFF state to the BCS state is
first order. Our theory does not provide information on the region,
extremely close to the critical point, denoted by the question mark in the
figure. Further calculations are needed to understand if the $s$-wave
superconducting state extends to the nonmagnetic side or it terminates on
the non-Fermi liquid phase.

In this paper we proposed the existence of an $s$-wave superconducting state
in weak ferromagnetic metals. We also propose a possible phase diagram of
weak ferromagnetic metals. These conclusions are based on the properties of
a weak ferromagnetic Fermi liquid. In principle this state should be
observable in pressure reduced critical temperature experiments similar to
these performed by the Cambridge group\cite{Lonzarich}.

We would like to thank A. Balatsky, J. Smith, A. Bishop and S. Trugman for
the useful suggestions. One of us (K.S.B.) would like to thank M.
Gul\'{a}csi for his hospitality. This work is sponsored in part by DOE Grant 
$DEFG0297ER45636$.

\end{document}